# Entropy Engineering-Regulated Electron-Phonon Coupling for Highly Efficient Photoluminescence in Se-doped WS$_2$

Chi Zhang[1,§], Quan Shen[4,§], Mengmeng Zhang[1,§], Zhiming Deng[5], Taishen Wu[1], Xuying Zhong[3], Gang Ouyang[3], Dongsheng Tang[3], Qi Zheng[1,2,†], Jiansheng Dong[4,†], Weichang Zhou[3,†]

[1] School of Physics and Electronics, Hunan University, Changsha 410082, People's Republic of China

[2] Greater Bay Area Innovation Institute, Hunan University, Guangzhou 511300, China, People's Republic of China

[3] School of Physics and Electronics, Key Laboratory of Low-dimensional Quantum Structures and Quantum Control of Ministry of Education, Key Laboratory for Multifunctional Ionic Electronic Materials and Devices of Hunan Normal University, Hunan Province Fundamental Research Center for Quantum Effects and Quantum Technology, Hunan Normal University, Changsha 410081, People's Republic of China

[4] Department of Physics, Jishou University, Jishou 416000, Hunan, China

[5] Key Laboratory of Hunan Province on Information Photonics and Freespace Optical Communications, School of physics and electrical sciences, Hunan Institute of Science and Technology, Yueyang, People's Republic of China

[§]These authors contributed equally to this work.
[†]Correspondence and requests for materials should be addressed to, Qi Zheng (e-mail: zhengqi@hnu.edu.cn), Jiansheng Dong (jsdong@jsu.edu.cn), Weichang Zhou (wchangzhou@hunnu.edu.cn).

**The limited quantum yield of strained monolayer transition metal dichalcogenides grown by vapor-phase methods and during transfer-based stacking poses a fundamental challenge for**

**their optoelectronic applications. Here, we introduce the concept of "entropy engineering" as a transformative strategy to selectively enhance light–matter interactions through controlled electron–phonon coupling. We unveil how tailored entropy introduced via precise selenium doping or interfacial van der Waals proximity can significantly amplify radiative recombination from momentum-dark excitons in WS$_2$ monolayers. Notably, we discover that slight selenium doping drastically enhances the photoluminescence (PL) of WS$_2$ under strain. While both undoped and heavily doped WS$_2$ suffer from strong PL quenching owing to the direct-to-indirect bandgap transition, lightly Se-doped samples exhibit an order-of-magnitude increase in emission intensity. This counterintuitive boost is traced to doping-induced structural disorder, which intensifies electron–phonon interactions and unlocks efficient phonon-assisted emission from otherwise non-radiative indirect excitons. Moreover, we demonstrate that van der Waals coupling to adjacent Se-doped layers can impart interfacial entropy and further augment PL via proximity effects. Our work highlights entropy engineering via controlled doping as a powerful strategy for activating high-efficiency light emission in atomically thin semiconductors.**

Two-dimensional transition metal dichalcogenides (TMDs) have attracted considerable research interest, both for fundamental studies of novel physical phenomena and for applications ranging from nanoelectronics and nanophotonics to sensing and actuation at the nanoscale[1–6]. In particular, monolayer semiconductor TMDs exhibit a direct bandgap and strong light–matter interactions, offering great potential for photonics and optoelectronics[2–6]. Nevertheless, despite advances in vapor-phase deposition for large-area synthesis[7–10], these monolayers often suffer from non-ideal photoluminescence (PL) intensity and quality. Key limiting factors include their atomic thickness, high defect density, strain, and structural disorder. Notably, strain—introduced during either growth or subsequent transfer and stacking processes[11]—can trigger a direct-to-indirect bandgap transition[12,13]. Semiconductors with an indirect bandgap usually leading to strong PL quenching and low quantum yield because electrons and holes located in different valleys. Intervalley excitons are momentum-dark since photons cannot provide the required momentum necessary for an indirect recombination[14–20]. Furthermore, dark excitons are highly interesting for TMD research, as they can lie energetically below bright excitons and hence have a significant impact on non-equilibrium dynamics as well as optical response of these materials[15–20]. Consequently, enhancing the quantum efficiency of momentum-dark intervalley excitons in strain-induced indirect-gap TMDs thus represents a critical challenge for their practical use in high-performance photonics and optoelectronics.

Here, we report an unconventional and counterintuitive strategy based on entropy engineering—using slight selenium (Se) doping and interfacial van der Waals proximity in strained $WS_2$ monolayers—to enhance electron–phonon coupling, leading to an orders-of-magnitude increase in PL radiation. Previous studies[21–26] have primarily focused on modulating the Se doping ratio in $WS_{2(1-x)}Se_{2x}$ alloys to achieve composition-tunable electronic and optical properties, while the effect of structural disorder induced by Se doping on electron–phonon coupling has remained unexplored. In this work, we systematically compare the PL characteristics of pure $WS_2$, lightly

Se-doped WS$_2$ (Se-WS$_2$), and heavily Se-doped WS$_2$ (alloyed). Under strain, both undoped WS$_2$ and WS$_{2(1-x)}$Se$_{2x}$ alloys exhibit severely suppressed quantum efficiency due to direct-to-indirect bandgap transition. In contrast, although Se-WS$_2$ also transitions to an indirect bandgap under strain, it exhibits strong electron–phonon coupling, which significantly enhances radiative recombination from momentum-dark indirect excitons via phonon-assisted intervalley charge transfer and multi-phonon participation. Our findings reveal that the radiative recombination of indirect excitons is highly susceptible to entropy modulation. The disorder introduced by light Se doping in both phonon and electronic structures facilitates strong electron–phonon coupling, enabling tunable enhancement of indirect exciton radiation by orders of magnitude. Furthermore, we observe that a van der Waals heterostructure composed of Se-WS$_2$ and pure WS$_2$ exhibits a pronounced proximity-enhanced luminescence at the interface. This effect unveils a novel mechanism whereby interlayer van der Waals contact modulates system disorder, thereby enhancing electron–phonon coupling and indirect exciton radiation. Our results not only demonstrate a new approach to enhance the light emission efficiency in TMDs but also highlight the profound role of disorder in mediating electron–phonon interactions, offering fresh physical insights and motivating new theoretical explorations.

The Se-WS$_2$ flakes were synthesized on Si/SiO$_2$ substrates via conventional two-step atmospheric-pressure chemical vapor deposition (CVD) with a rapid cooling process employed to introduce tensile strain into the as-synthesized material (see Fig. S1 and Methods for details). Obviously, such a method generally does not allow precise control over the doping concentration of Se or its spatial distribution. In our experiments, triangular flakes are typically observed, wherein the Se-WS$_2$ region is situated at the center of the triangle, surrounded by intrinsic WS$_2$ along the outer region, thereby forming an in-plane lateral heterostructure (Fig. 1a). Such structure indicate that the growth process initially forms small WSe$_2$ domains, and the subsequent introduction of sulfur replaces the selenium in the existing WSe$_2$ domains (sulfurization), while simultaneously promoting the lateral epitaxial growth of WS$_2$ along the edges of the domains. Figure 1b shows two typical Raman spectra obtained from the pure WS$_2$ and the Se-WS$_2$ regions under 532 nm excitation, represented by the black and red curves respectively, both exhibiting distinct characteristic peaks of intrinsic WS$_2$ monolayer (See Fig. S2 and S3 for more experimental results and detail discussion). The Raman spectra exhibit resonance features, and apart from peaks due to the Brillouin zone center optical phonons, various second-order peaks emerge throughout the 100−450 cm$^{-1}$ frequency range (See Fig. S4 for detail discussion of various Raman modes). The most prominent modes in both spectra exhibit the following shared features: a double-resonance Raman scattering peak involving two longitudinal acoustic phonons (2LA) at 350 cm$^{-1}$, and zone-center optical phonons corresponding to in-plane (E$_{2g}$) and out-of-plane (A$_{1g}$) vibrations at 354 cm$^{-1}$ and 418 cm$^{-1}$, respectively. Mechanical strain alters interatomic distances, thereby shifting phonon frequencies through modifications in lattice dynamics, which can be monitored via Raman spectroscopy. Generally, the A$_{1g}$ mode exhibits a weak response to in-plane strain, whereas the E$_{2g}$ mode shows a more pronounced shift due to its vibrational direction being within the plane where strain is applied. In previous studies[27–31], the strain magnitude has often been quantified using the frequency separation (Δω) between the $E_{2g}$ and $A_{1g}$ peaks. In our as-grown $WS_2$ flakes, Δω is measured to be approximately 64 cm$^{-1}$, compared to a reported value of ~62 cm$^{-1}$ for transferred $WS_2$ where strain is released[27,28]. Based on this difference, the residual strain in our as-grown lateral heterostructure is globally estimated to be around 0.5% to 1%. In addition to strain information, Raman spectroscopy can also reveal doping characteristics. The primary spectral difference between the two regions lies in the emergence of two characteristic peaks at approximately 376 cm$^{-1}$ and 396 cm$^{-1}$ (Fig. 1b), which are exclusively observed in the Se-$WS_2$ region. These peaks are attributed to disorder-activated vibrational modes, tentatively assigned to in-plane ($E_{2g}$) and out-of-plane ($A_{1g}$) vibrations of S–W–Se[21,32,33] (See Fig. S4 for detail discussion).

These Raman characterization results are consistent with the formation of a lateral $WS_2$/Se-$WS_2$ heterostructure, as schematically illustrated in Fig. 1a. To further evaluate the Se doping concentration and its influence on the band structure and excitonic properties, we performed PL measurements (Fig. 1c). Both the Se-$WS_2$ and pure $WS_2$ regions exhibit PL emission at nearly identical energies, comparable to previously reported[27–31] peak positions for intrinsic $WS_2$. This suggests a low Se doping concentration in the Se-$WS_2$ region, estimated to be less than 5% (We aimed to estimate a conservative upper limit; the actual doping level is likely even lower.). Moreover, the PL signal from the pure $WS_2$ region is notably weak, comparable in intensity to that of the bilayer region, and exhibits a double-peak feature with energy difference $\Delta E \approx 16$ meV (additional PL results are provided in Fig. S2c). These observations suggest that the applied strain induces a transition from direct to indirect bandgap in WS$_2$, leading to strong PL quenching. To validate this hypothesis, we conducted first-principles calculations. As shown in Fig. 1d and 1e, both pure WS$_2$ and Se-WS$_2$ (Se doping concentration of 2.08%) undergo a direct-to-indirect bandgap transition under biaxial strain of 1% (see Figs. S5 and S6 for details and discussion of the first-principles calculations). Similar to bilayer TMDs[34–38], as shown in Fig. 1f, photoexcitation initially creates direct excitons at the K-points. These subsequently undergo rapid phonon-assisted relaxation to the lower-energy indirect band edges, forming momentum-mismatched indirect excitons. Radiative recombination of these indirect excitons requires a slow, phonon-mediated second-order process to conserve momentum, resulting in low quantum efficiency. Nevertheless, thermal population of the higher-energy direct states can also produce a weaker direct emission, giving rise to a higher-energy emission peak alongside the dominant low-energy indirect peak, thereby resulting in a characteristic dual-peak PL spectrum. Remarkably, however, under the same strain conditions, the Se-WS$_2$ region exhibits highly efficient PL—showing an order-of-magnitude enhancement compared to the undoped region (See Fig. S3 for more experimental results of additional sample)—despite undergoing a similar direct-to-indirect bandgap transition.

To investigate the underlying mechanism of Se-WS$_2$ maintains efficient luminescence under strain, we performed Raman and PL measurements on another sample grown by the same method. Figure 2a shows an optical image of a composition-graded WS$_{2(1-x)}$Se$_{2x}$ monolayer, which exhibits a well-faceted regular triangular shape with a distinct color gradation from the center to the edge. Both Raman and PL spectra, as functions of spatial position, clearly demonstrate the composition gradient within the sample: pure WS$_2$ is located at the outermost edge, while the Se doping concentration gradually increases toward the center (Figs. 2b and c). For clarity, the sample can be divided into three distinct regions: the outermost region consisting of pure WS$_2$, the central region with a higher Se doping concentration for WS$_{2(1-x)}$Se$_{2x}$ alloy, and an intermediate region exhibiting slight Se doping in WS$_2$. As shown in Fig. 2d, both pure WS$_2$ and the WS$_{2(1-x)}$Se$_{2x}$ alloy exhibit weak luminescence intensity due to the strain-induced direct-to-indirect bandgap transition (see Fig. S7 for the first-principles calculations of WS$_{2(1-x)}$Se$_{2x}$ alloy), whereas the region with slight Se-doped WS$_2$ demonstrates highly efficient PL intensity. A sublinear relationship widely used for ternary alloy semiconductors[33]: $E_g(x) = (1-x)E_g(WS_2) + xE_g(WSe_2) - bx(1-x)$, where $x$ is the composition ratio of Se/(S + Se) and $b$ is the bowing factor (The value of b is given as 0.154[33], $E_g(WS_2) = 1.975\ eV$, $E_g(WSe_2) = 1.630\ eV$). It can be estimated that a significant enhancement in PL intensity occurs only when the doping concentration is less than 15% (with a predicted $E_g(x)$ lower limit of 1.900 eV, see Fig. S4 for detail discussion).

With increasing Se doping concentration, not only does the PL intensity exhibit a non-monotonic variation, but the full width at half maximum (FWHM) of the PL peak also follows a non-monotonic trend. As illustrated in Fig. S8a, a comparative statistical analysis further reveals the spatial variations in both PL intensity and FWHM. It can be observed that although the FWHM similarly displays a non-monotonic dependence on Se concentration, its behavior is not entirely identical to that of the PL intensity. Notably, the FWHM reaches its maximum only in regions with relatively higher doping (approximately 7%–15%). In contrast, at lower doping levels (<7%), the FWHM remains nearly identical to that of pure WS$_2$ (Fig. 1c and Fig. S8b). Figure 2e compares the PL spectra of pure WS$_2$, Se-WS$_2$ (around 7%–15%), and WS$_{2(1-x)}$Se$_{2x}$ alloy. Clearly, the Se-WS$_2$ (around 7%–15%) exhibits a broad FWHM and an asymmetric lineshape. On the low-energy side of this asymmetric spectrum, clearly resolved multi-peak features with regular energy spacing are observed, ruling out contributions from charged excitons or biexcitons. Reasonably, we attribute this spectral feature to phonon replica features associated with indirect exciton radiation involving multiple phonons.

To further elucidate the origin of the multi-peak features and asymmetric lineshape in the PL spectrum of Se-WS$_2$ (~11%), we performed power-dependent PL measurements. As shown in Fig. 3a, at low laser power, the spectrum is dominated by an asymmetric profile. With increasing laser power, the multi-peak features become more pronounced and exhibit equienergy phonon replicas with $\Delta E \approx 17$ meV (corresponding to the out-of-plane acoustic phonon (ZA) mode, see Fig. S4 for detail discussion). Furthermore, the intensity of the lower-energy peaks increases more rapidly with higher excitation power. This behavior can be attributed to the increased photoexcited carrier density at higher laser powers, which enhances the electron–phonon interaction strength[39,40]. At higher power densities, the increased carrier density raises the probability of collisions with scattering centers. As a result, more carriers undergo relaxation via the emission of one or multiple phonons. This relaxation process populates the lower-energy states more efficiently. Consequently, the indirect radiative recombination process involves the emission of multiple phonons per electron–hole pair (Fig. 3b). Therefore, at low Se doping concentrations (<7%), the process of indirect radiative recombination in WS$_2$ is activated. In this regime, disorder provides an additional center-of-mass momentum that enables the emission of momentum-dark excitons, with each electron–hole pair recombining radiatively with the assistance of a single phonon. As the doping level increases (7%–15%), the number of phonons involved in the recombination process rises, leading to multi-phonon participation in the radiative recombination of each electron–hole pair. When the doping concentration exceeds 15%, the system becomes fully alloyed, structural disorder is reduced, and the pathway for indirect radiative recombination is suppressed.

To investigate the physical mechanism by which Se doping modulates the electron–phonon coupling in WS$_2$, we performed first-principles calculations to compute the electronic density of states (DOS) at the Fermi surface (FS) and the phonon spectra of pure WS$_2$ and the Se-WS$_2$ (2.08%) under 1% biaxial strain (Figs. 3c-f). The electron-phonon coupling strength can be described by $\lambda = \sum_{ph} \frac{\langle g^2 \rangle_F}{N_f \hbar \omega_{ph}}$, where $N_f$ and $\hbar \omega_{ph}$ are the DOS at the FS and the phonon energy, $\sum_{ph}$ and $\langle g^2 \rangle_F$ are the sum among different phonon modes and the square sum of electron-phonon coupling matrix elements on FS, respectively[41,42]. A comparative analysis of the DOS at the FS near the valence band maximum (VBM) reveals a suppression in the Se-WS$_2$ relative to pure WS$_2$ under 1% biaxial strain (Figs. 3c and d), suggestive of enhanced electron-phonon coupling. This interpretation is corroborated by the phonon spectra (See Fig. S9 for more results), which reveal reduced phonon group velocity in Se-WS$_2$ compared to pure WS$_2$ and WS$_{2(1-x)}$Se$_{2x}$ alloy ($x = 33\%$). These results further support the demonstration that slightly Se doping significantly enhances disorder and strengthens electron-phonon coupling.

More remarkably, we discovered that van der Waals contact between Se-WS$_2$ and pure WS$_2$ can likewise introduce disorder at the interface, enhancing electron–phonon coupling and yielding highly efficient PL. As simulated in Fig. 4a, using a similar growth method—only reversing the introduction sequence of the Se and S sources (S first followed by Se)—we successfully fabricated the WS$_2$/Se-doped WS$_2$ vertical heterostructures. Figure 4b presents a typical optical micrograph of the vertically stacked heterostructure, featuring a regular triangular shape with multilayer AA stacking. As expected, the additional Raman peaks induced by Se doping were only observed in the centrally located multilayer stacked region of the sample, whereas the outer monolayer area exhibited the characteristic Raman spectra of pure WS$_2$ (Fig. 4c). Similarly, the PL signal from the monolayer WS$_2$ is notably weak (black curve), comparable in intensity to that of the bilayer region (blue curve), suggesting a direct to indirect bandgap transition under strain. In contrast, at the interface between the monolayer and bilayer regions (on the monolayer side), the PL intensity exhibits a significant enhancement—more than 50 times greater than that in other areas of the monolayer (Figs. 4d and e).

To rule out the possibility that the highly efficient PL originates from direct Se doping in the central monolayer WS$_2$ region—which would form a lateral heterostructure similar to those shown in Figs. S2 and S3—we fabricated bilayer heterostructures using the S-source-first followed by Se-source growth sequence, as presented in Figs. S10 and S11. In both the AB-stacked bilayer heterostructure (Fig. S10) and the AA-stacked bilayer with an irregularly shaped top layer (Fig. S11), the PL enhancement is exclusively localized at the interface between the monolayer and bilayer regions. These results provide compelling evidence that the emission enhancement in monolayer $WS_2$ is precisely dictated by the morphology of the overlying Se-$WS_2$ layer and occurs solely at the interfacial contact area. Therefore, we can conclusively exclude direct Se doping of the monolayer $WS_2$ as the origin of the PL enhancement. Furthermore, we synthesized a specially designed vertically stacked heterostructure (Fig. S12), in which the bottom layer consists of pure $WS_2$ and the top layer is a lateral heterostructure of pristine $WS_2$ and Se-$WS_2$. Significant PL enhancement in the monolayer region was observed only at the three corners contacting the Se-$WS_2$ top layer, while no enhancement occurred at interfaces with the pristine $WS_2$ part of the top layer. This further confirms that the luminescence enhancement stems from the van der Waals contact between Se-$WS_2$ and monolayer $WS_2$, where proximity-induced interfacial disorder and enhanced electron–phonon coupling play critical roles. Moreover, we found that the proximity-induced interfacial PL enhancement via van der Waals contact is not limited to the case where the upper layer is slightly Se-doped $WS_2$. As shown in Fig. S12, the PL peak of the top Se-$WS_2$ layer is located at 1.84 eV, corresponding to an estimated doping level of approximately 30%. Remarkably, even such an alloyed $WS_{2(1-x)}Se_{2x}$ layer can significantly enhance the PL of the underlying monolayer through proximity coupling, thereby overcoming the limitations associated with direct doping.

Our results demonstrate that slight Se doping effectively enhances the PL in $WS_2$ monolayers by strengthening the coupling between electrons and ZA phonons via entropy-mediated disorder engineering. Compared to both pristine $WS_2$ and fully alloyed $WS_{2(1-x)}Se_{2x}$, the slightly Se-doped $WS_2$ exhibits significantly enhanced disorder, which directly reduces the DOS near the FS and induces pronounced phonon flattening alongside an increased phonon DOS. These changes collectively lead to a marked enhancement in electron–phonon coupling, facilitating phonon-assisted radiative recombination of indirect excitons. The tunability of disorder via Se concentration underscores the role of entropy engineering in low-dimensional systems: unlike high-entropy multicomponent alloys[43–45], even minor single-element doping can effectively introduce configurational disorder and modify the vibrational and electronic properties. This approach provides a new pathway for tailoring light–matter interactions in atomically thin semiconductors through entropy-regulated electron–phonon coupling.

**Data availability**

All data supporting the findings of this study are available from the corresponding author upon request.

**Acknowledgments:**

This work was supported by the National Natural Science Foundation of China (Grant Nos. 12404204, 12364007, 62575100), Natural Science Foundation of Hunan Province (Grant Nos. 2024JJ6116), Guangdong Basic and Applied Basic Research Foundation (Grant Nos. 2025A1515010434), Natural Science Foundation of Changsha City (Grant Nos. kq2402049, kq2402151).

**Author contributions statement**

Q.Z. and W.Z designed the experiment. Q.Z., C.Z. and M.Z. performed the sample synthesis, characterization and optical measurements. Q.S. carried out the theoretical calculations. J.D. conceived and provided advice on the theoretical calculations. Q.Z., J.D. and C.Z. analyzed the data and wrote the paper. All authors participated in the data discussion.

**Competing interests statement**

The authors declare no competing interests.

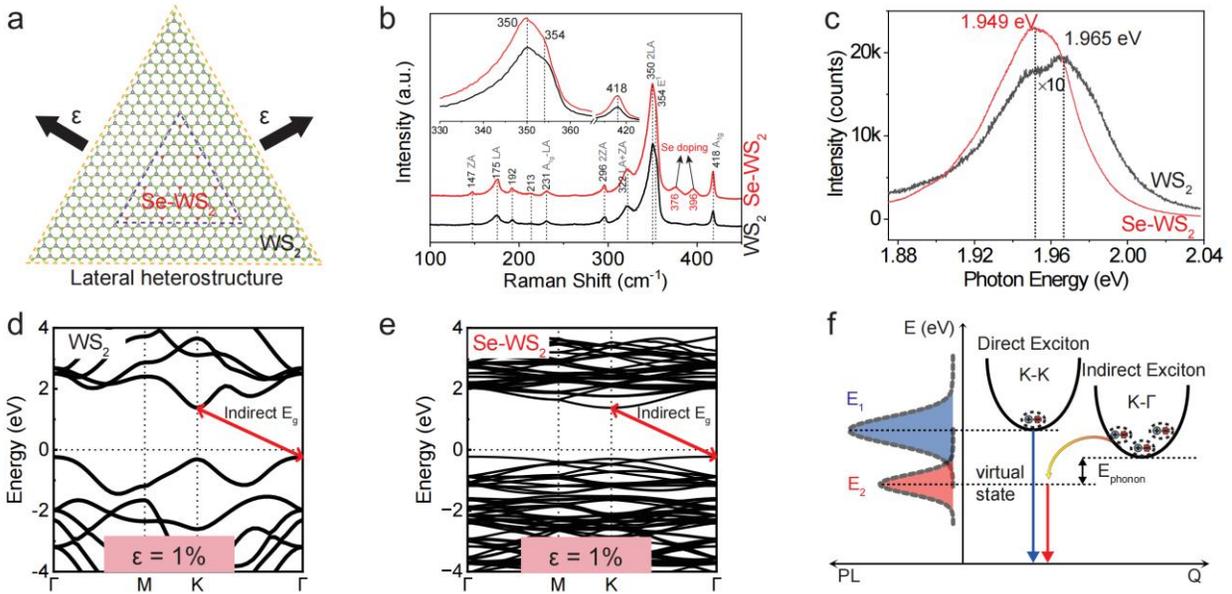

**Fig. 1 | Optical characterization and energy band structure in WS$_2$/Se-WS$_2$ lateral heterostructure. a,** Schematic diagram of a monolayer WS$_2$/Se-WS$_2$ lateral heterostructure under biaxial strain ε. The yellow and purple dashed boxes indicate the contour of the whole WS$_2$ monolayer and the interface of the WS$_2$/Se-WS$_2$ lateral heterostructure, respectively. **b,** Raman spectra obtained from pure WS$_2$ (marked by black star in Fig. S2a), and Se-WS$_2$ (marked by red star in Fig. S2a), respectively. Inset: Zoom-in of the Raman spectra in the range from 330 to 423 cm$^{-1}$. **c,** PL spectra obtained from pure WS$_2$ (marked by black star in Fig. S2a), and Se-WS$_2$ (marked by red star in Fig. S2a), respectively. **d** and **e**, the calculated electronic band structures of WS$_2$ and the Se-WS$_2$ (Se doping concentration of 2.08%) monolayer under application of biaxial strain ($\varepsilon = 1\%$). The application of strain induces a direct-to-indirect bandgap transition in both WS$_2$ and Se-WS$_2$ monolayer. **f,** Sketch of direct and indirect decay channels for excitons showing the underlying scattering processes in the excitonic center-of mass dispersion (right) and the corresponding PL signals (left).

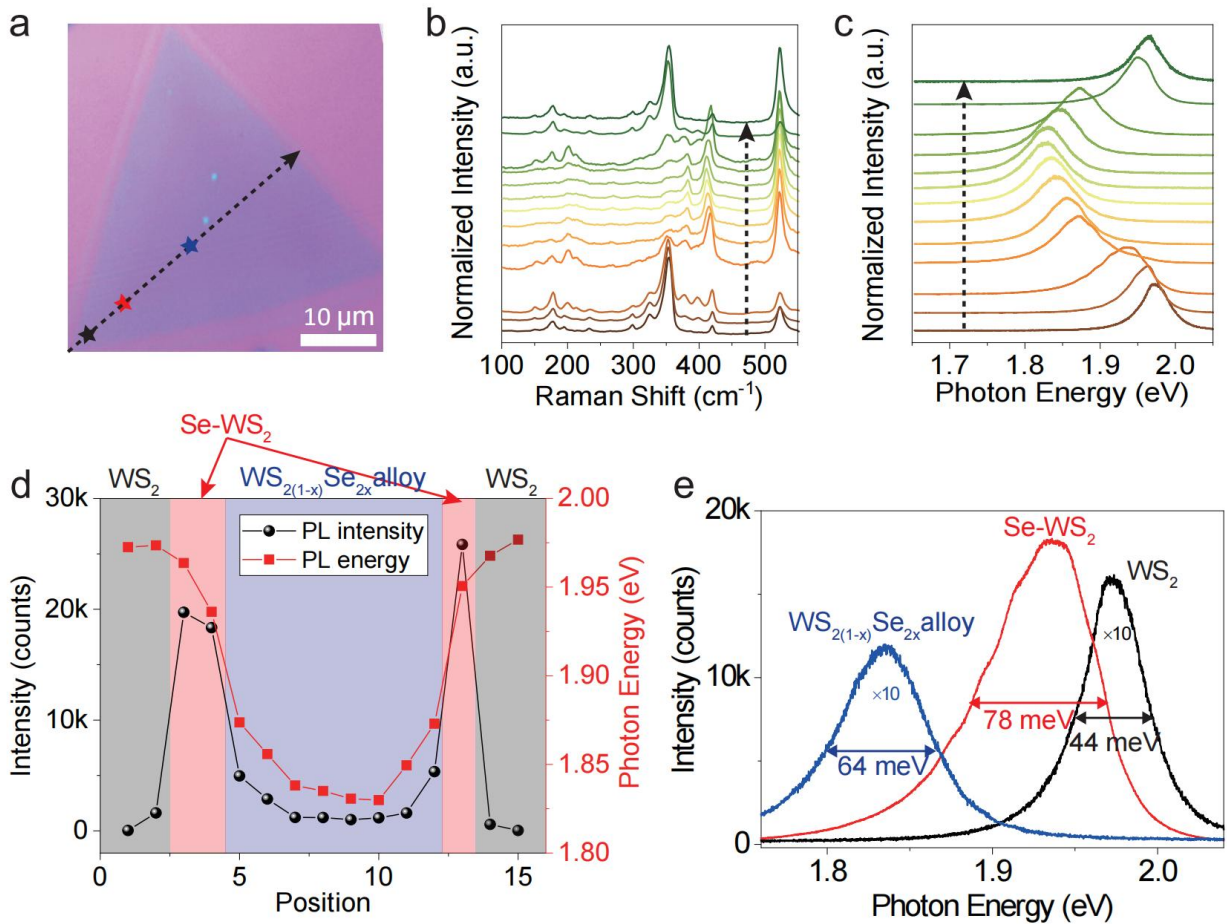

**Fig. 2 | Raman and PL characterization of composition-graded WS$_{2(1-x)}$Se$_{2x}$ monolayer. a,** Optical micrographs of a representative composition-graded WS$_{2(1-x)}$Se$_{2x}$ monolayer, showing a progressive color gradient from the center to the edge. **b** and **c**, Raman and normalized PL spectra measured along the dashed arrow (marked in panel a). **d**, Spatial dependence of the PL emission intensity (black dots) and energy (red dots), obtained from panel c. Based on this evolution pattern, the flake can be divided into three distinct compositional regions: pure WS$_2$ (black shaded area), Se-WS$_2$ (Se doping level below 15%), and the alloyed WS$_{2(1-x)}$Se$_{2x}$ (Se doping level above 15%). The Se-WS$_2$ region exhibits a significant PL enhancement compared to both the pure WS$_2$ and the alloyed WS$_{2(1-x)}$Se$_{2x}$ regions. **e**, Representative PL spectra acquired from three distinct

compositional regions, showing significantly different FWHM: pure $WS_2$ (~44 meV), Se-$WS_2$ (~78 meV), and alloyed $WS_{2(1-x)}Se_{2x}$ (~64 meV).

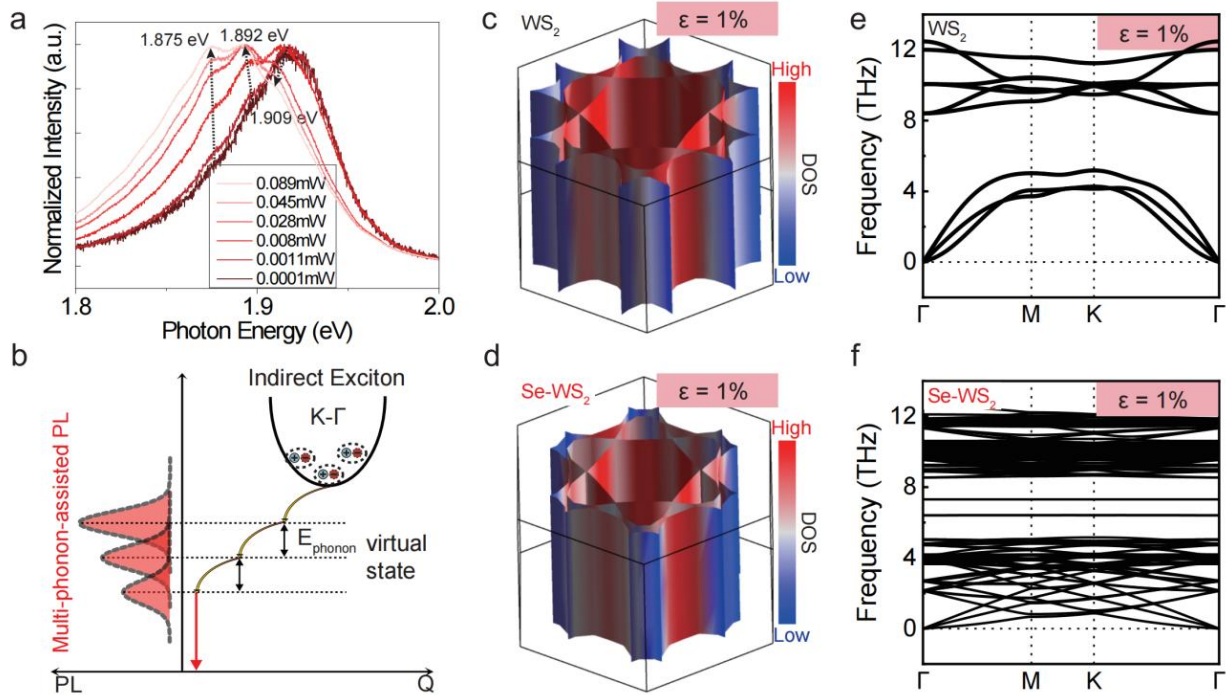

**Fig. 3 | Multi-phonon-assisted indirect band exciton radiation and strong electron-phonon coupling mechanism. a,** Laser power-dependent PL spectra measured at the Se-WS$_2$ (Se doping concentration ~11%). **b,** Sketch of the multi-phonon-assisted indirect exciton radiative decay channels, illustrating the underlying scattering processes in the excitonic center-of-mass dispersion (right) and the corresponding PL spectra (left). **c** and **d,** The computed electronic density of states near the FS at the VBM for both WS$_2$ and the Se-WS$_2$ monolayers (Se doping concentration 2.08%) under 1% biaxial strain, as obtained from first-principles calculations. **e** and **f,** The computed phonon spectra for both WS$_2$ and the Se-WS$_2$ monolayers (Se doping concentration 2.08%) under 1% biaxial strain.

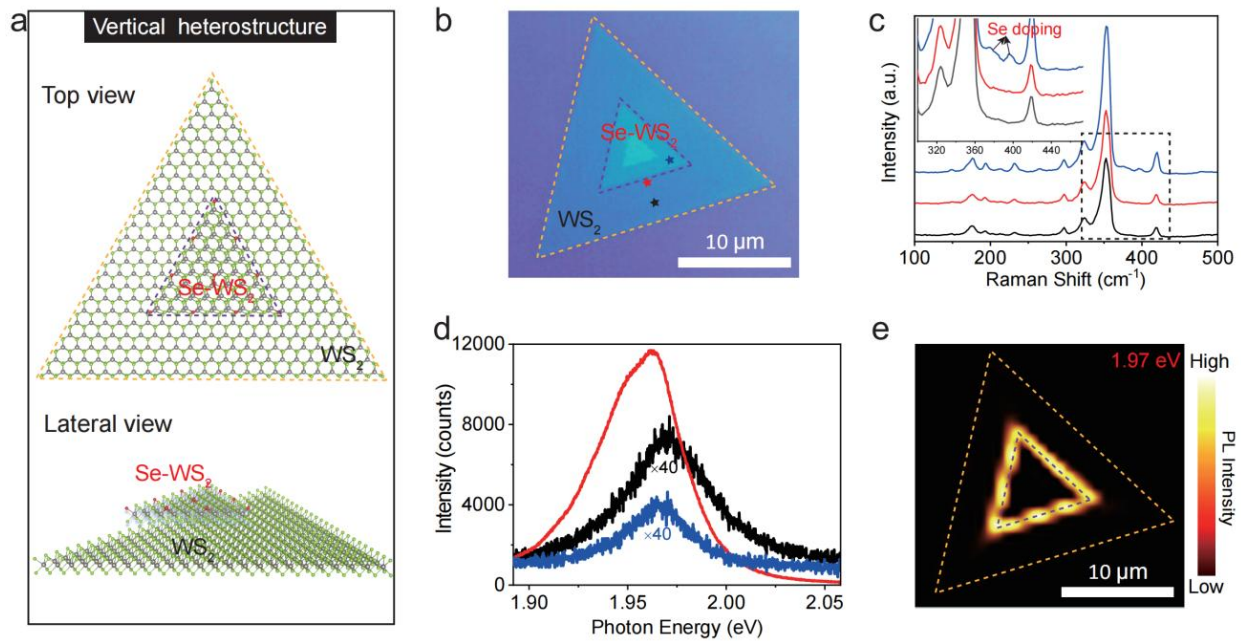

**Fig. 4 | Optical characterization of WS$_2$/Se-WS$_2$ vertical heterostructure. a,** Schematic diagram of a bilayer WS$_2$/Se-WS$_2$ vertical heterostructure (top and side views). The yellow and purple dashed boxes outline the bottom-layer WS$_2$ and the top-layer Se-WS$_2$, respectively. **b,** The optical image of a trilayer WS$_2$ vertical heterostructure sample. The yellow and purple dashed boxes outline the bottom-layer WS$_2$ and the middle-layer Se-WS$_2$, respectively. **c** and **d**, Raman and PL spectra obtained from bottom-layer WS$_2$ (black curve corresponds to the black star in panel b), monolayer–bilayer interfacial boundary (red curve corresponds to the red star in panel b), and middle-layer Se-WS$_2$ (blue curve corresponds to the blue star in panel b) regions, respectively. Inset in panel c: Zoom-in of the Raman spectra marked by the black dashed box. **e,** Fixed-energy PL spatial distribution map at 1.97 eV.